\documentclass[aps,floatfix,showpacs,preprint,preprintnumbers,superscriptaddress,amsmath,amssymb,pra]{revtex4-1}

\usepackage{graphicx}
\begin{document}

\preprint{}

\title{\boldmath Anomalous quantum-critical spin dynamics in YFe$_2$Al$_{10}$}

\author{K. Huang}
\altaffiliation{Current address: National High Magnetic Field Laboratory, Florida State University, 1800 E. Paul Dirac Dr., Tallahassee, Florida 32310, USA.}
\author{C. Tan}
\author{J. Zhang}
\author{Z. Ding}
\affiliation{State Key Laboratory of Surface Physics, Department of Physics, Fudan University, Shanghai 200433, China}
\author{D.~E. MacLaughlin}
\affiliation{Department of Physics and Astronomy, University of California, Riverside, California 92521, USA}
\author{O.~O. Bernal}
\affiliation{Department of Physics and Astronomy, California State University, Los Angeles, California 90032, USA}
\author{P.-C. Ho}
\affiliation{Department of Physics, California State University Fresno, Fresno, California 93740, USA}
\author{C. Baines}
\affiliation{Laboratory for Muon-Spin Spectroscopy, Paul Sherrer Institute, CH-5232 Villigen, Switzerland}
\author{L.~S. Wu}
\author{M.~C. Aronson}
\altaffiliation{Current address: Department of Physics \& Astronomy, Texas A\&M University, College Station, Texas 77843, USA.}
\affiliation{Department of Physics and Astronomy, Stony Brook University, Stony Brook, New York 11794, USA}
\affiliation{Condensed Matter Physics and Materials Science Department, Brookhaven National Laboratory, Upton, New York 11973, USA.}
\author{L. Shu}
\altaffiliation[Corresponding Author: ]{leishu@fudan.edu.cn.}
\affiliation{State Key Laboratory of Surface Physics, Department of Physics, Fudan University, Shanghai 200433, China}
\affiliation{Collaborative Innovation Center of Advanced Microstructures, Nanjing 210093, China}

\date{\today}

\hyphenation{TRIUMF}

\begin{abstract}
We report results of a muon spin relaxation ($\mu$SR) study of YFe$_2$Al$_{10}$, a quasi-2D nearly-ferromagnetic metal in which unconventional quantum critical behavior is observed. No static Fe$^{2+}$ magnetism, with or without long-range order, is found down to 19~mK\@. The dynamic muon spin relaxation rate~$\lambda$ exhibits power-law divergences in temperature and magnetic field, the latter for fields that are too weak to affect the electronic spin dynamics directly. We attribute this to the proportionality of $\lambda(\omega_\mu,T)$ to the dynamic structure factor~$S(\omega_\mu,T)$, where $\omega_\mu \approx 10^5$--$10^7~\mathrm{s}^{-1}$ is the muon Zeeman frequency. These results suggest critical divergences of $S(\omega_\mu,T)$ in both temperature and frequency. Power-law scaling and a 2D dissipative quantum XY (2D-DQXY) model both yield forms for $S(\omega,T)$ that agree with neutron scattering data ($\omega \approx 10^{12}~\mathrm{s}^{-1}$). Extrapolation to $\mu$SR frequencies agrees semi-quantitatively with the observed temperature dependence of $\lambda(\omega_\mu,T)$, but predicts frequency independence for $\omega_\mu \ll T$ in extreme disagreement with experiment. We conclude that the quantum critical spin dynamics of YFe$_2$Al$_{10}$ are not well understood at low frequencies.

\end{abstract}


\maketitle


\section{INTRODUCTION}

Quantum phase transitions (QPTs) occur at absolute zero temperature, and are fundamentally different phenomena than thermally-driven transitions. Ferromagnetic QPTs in metals, in particular, exhibit a broad spectrum of properties~\cite{[{For a recent review, see }] BRANDO16}. Theoretical work suggests that clean quantum-critical ferromagnets should exhibit a discontinuous first-order phase transition, but it has been argued that inclusion of disorder~\cite{BKV05,LOHNEYSEN07} or strong quantum fluctuations in low-dimensional systems~\cite{BKV05} can destroy the first-order character~\cite{WU14}, resulting in a quantum critical point. Thus investigating ferromagnetic quantum critical points (QCPs) in clean systems is crucial but difficult, as genuine reduced dimensionality and elimination of disorder are both hard to achieve. Furthermore, many measurement techniques (e.g., conventional magnetization, NMR) require applied magnetic fields that can tune the system away from criticality.

First reported in 1998~\cite{TEJ98}, the layered compound YFe$_2$Al$_{10}$ is a rare example of a quasi-two-dimensional (2D) material that is on the threshold of ferromagnetism. Magnetic susceptibility and specific heat measurements reveal unusual divergences at low temperatures~\cite{STRYDOM10,STRYDOM11,PARK11,KHUNTIA12,WU14}, differing significantly from conventional Fermi-liquid metals. The divergences are quenched by magnetic field~\cite{PARK11}, which is characteristic of a ferromagnet. Such divergent or non-Fermi liquid behavior is thought to be due to quantum fluctuations associated with a QCP~\cite{MAPLE10}. YFe$_2$Al$_{10}$ is a clean~\cite{PARK11} stoichiometric compound, and is close to quantum criticality without tuning by chemical substitution, pressure, or magnetic field~\cite{PARK11,WU14}.

Recent neutron scattering experiments~\cite{GANNON17u} revealed no long-range order in YFe$_2$Al$_{10}$ (i.e., no divergence of the spatial correlation length), but indicated a divergence of the fluctuation time scale. This contrasts strongly with the usual paradigm for critical phenomena, in which both spatial and temporal correlation scales diverge at the transition. It agrees, however, with the separability of these correlations found in the 2D dissipative quantum XY (2D-DQXY) model proposed by Varma and collaborators~\cite{AJI10,ZHU15,HOU16}. The momentum-integrated magnetic dynamic structure factor~$S(\omega,T)$ was observed to be essentially temperature independent, and over the range of energies in the experiment (0.35--0.7~meV, $\omega \approx 5\text{--}10 \times 10^{11}\ \mathrm{s}^{-1}$) its frequency dependence could be fit equally well by either power-law scaling suggested by earlier work~\cite{WU14} or the 2D-DQXY functional form~\cite{[{}] [{; C.~M. Varma, unpublished.}] VARMA17u}. The absolute value of $S(\omega,T)$ was determined in this study.

Magnetic resonance techniques such as NMR and muon spin rotation/relaxation ($\mu$SR)~\cite{SCHENCK85,*BLUNDELL99,*BREWER03,*YAOUANC11} are local probes of magnetic behavior, and are therefore complementary to bulk measurements and reciprocal-space (scattering) probes. In magnetic resonance the dynamic or ``spin-lattice'' relaxation rate~$\lambda(\omega,T)$ of the spin probe (nucleus or muon) also measures $S(\omega,T)$~\cite{MORIYA63,*JACCARINO67,*CHAKRAVARTY90}, but at much lower frequencies ($10^5$--$10^7~\mathrm{s}^{-1}$) than neutron scattering. Absolute values of $S(\omega,T)$ can thus be determined at both neutron and $\mu$SR frequencies. An advantage of $\mu$SR compared to NMR is that no external magnetic field is needed, since the muons in the incident beam are 100\% spin polarized~\cite{SCHENCK85,*BLUNDELL99,*BREWER03,*YAOUANC11}. A field of any magnitude can be applied if desired.

This article reports $\mu$SR measurements of $\lambda(\omega,T)$ in single crystals of YFe$_2$Al$_{10}$. No evidence of magnetic order was found down to 19 mK\@. In contrast to a previous study~\cite{ADROJA13}, significantly enhanced dynamic muon relaxation was observed at low temperatures and magnetic fields. This strongly suggests the presence of quantum-critical spin fluctuations. The temperature dependence of $\lambda(\omega,T)$ agrees with that from extrapolation to low frequencies of $S(\omega,T)$ from either power-law scaling description or the 2D-DQXY model, although the divergence of $\lambda$ in temperature is cut off below $\sim$0.1~K; YFe$_2$Al$_{10}$ is close to but perhaps not exactly at quantum criticality. However, both of these approaches predict frequency independence of $S(\omega,T)$ at low frequencies, whereas a strong divergence ($\lambda \propto 1/H$ at 25~mK) is observed. This rather extreme disagreement leads us to conclude that the low-frequency quantum critical dynamics in this compound are not well understood, and that more work is required.

\section{EXPERIMENT}

Single crystals of YFe$_2$Al$_{10}$ were grown in an aluminum flux as described previously~\cite{PARK11}. Separate samples were prepared with $b$ and $c$ axes normal to the large faces. Oriented single crystals were mounted on silver sample holders using dilute GE varnish. Zero-field and longitudinal-field (applied field~$\mathbf{H}_L$ parallel to the initial muon spin polarization~$\mathbf{P}_\mu$) $\mu$SR experiments (ZF-$\mu$SR and LF-$\mu$SR, respectively) were carried out at the M15 and M20 beam lines at TRIUMF, Vancouver, Canada, and at the LTF beam line at the Paul Scherrer Institute, Villigen, Switzerland. The time-differential $\mu$SR technique~\cite{BREWER03,YAOUANC11} was used, in which the evolution of the ensemble muon spin polarization is monitored via measurements of the decay positron count rate asymmetry~$A(t)$ vs.\ time~$t$ after muon implantation.

\subsection{\boldmath Zero-field $\mu$SR} \label{par:ZF}

\begin{figure}[t]
 \includegraphics[clip=,width=0.45\textwidth]{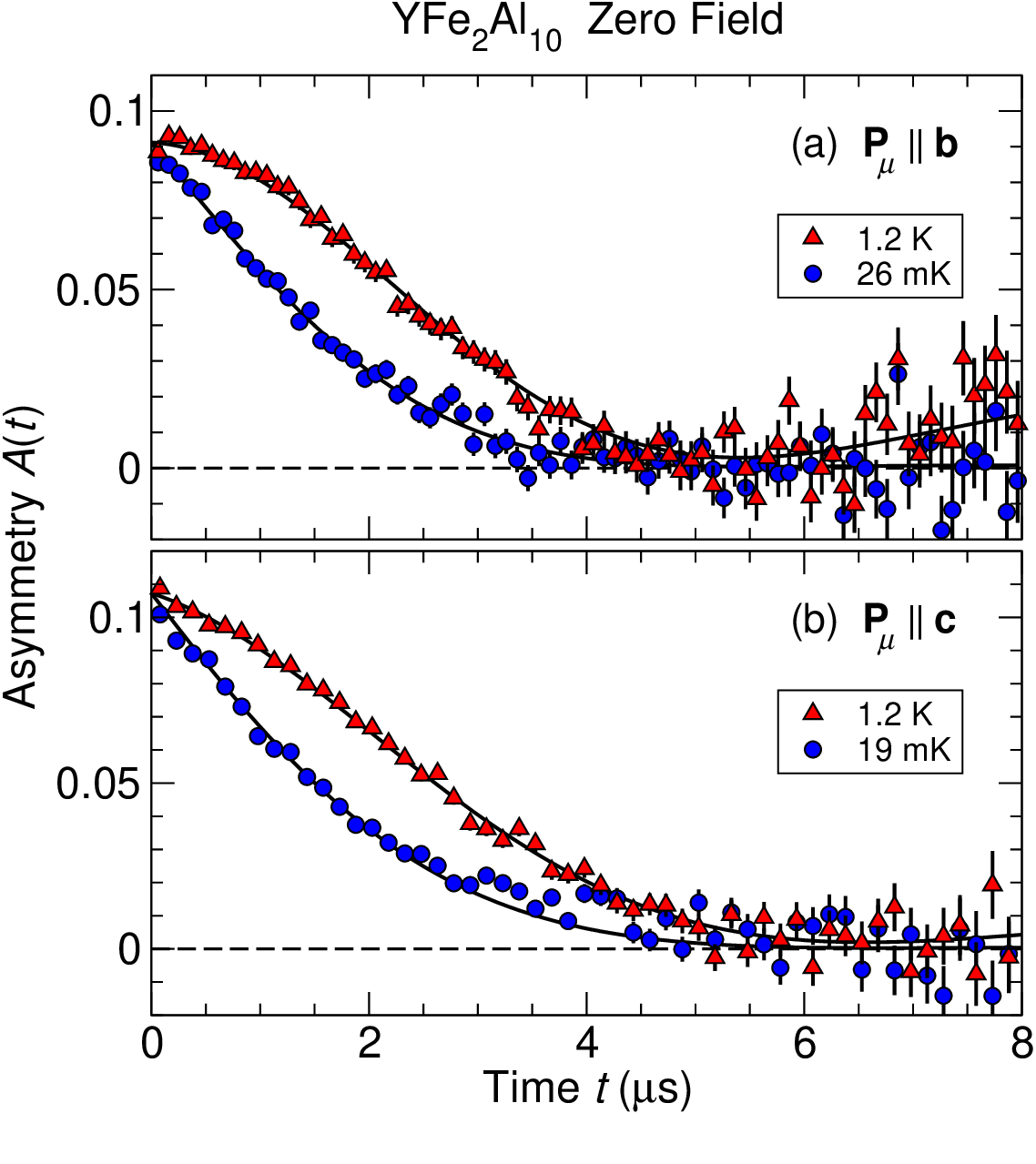}
 \caption{Zero-field (ZF) $\mu$SR asymmetry spectra (time evolution of decay positron count rate asymmetry) from single crystals of YFe$_{2}$Al$_{10}$. (a)~Initial muon spin polarization~$\mathbf{P}_\mu$ parallel to the $b$ crystal axis. (b): $\mathbf{P}_\mu\,{\parallel}\,\mathbf{c}$. Curves: fits of exponentially-damped ZF static Gaussian Kubo-Toyabe functions to the data (see text). No asymmetry loss or oscillations are observed, evidence that there is no static Fe$^{2+}$ magnetism down to 19~mK.}
 \label{fig1}
\end{figure}

ZF-$\mu$SR experiments were performed over the temperature range 19~mK--10~K\@. ZF-$\mu$SR asymmetry spectra are roughly temperature independent for $T \sim 1$--10~K, but exhibit a strong temperature dependence at lower temperatures. Representative spectra are shown in Fig.~\ref{fig1} for $\mathbf{P}_\mu$ parallel to the $b$ and $c$ crystal axes [Figs.~\ref{fig1}(a) and \ref{fig1}(b), respectively].
A constant background signal from muons that stop in the silver sample holder has been subtracted. We find no evidence, such as loss of initial asymmetry or oscillations due to precession in an internal field, for a transition to a state of static Fe$^{2+}$ magnetism, ordered or disordered, down to 19 mK.

The ZF spectra are well described by the functional form
\begin{equation}
\label{eq:ZF-asym}
A(t) = A_0\exp(-\lambda_\mathrm{ZF} t)G_\mathrm{ZF}^\mathrm{KT}(\Delta, t) \,,
\end{equation}
where $A_0$ is the initial count-rate asymmetry, and
\begin{equation}
\label{eq:KT}
G_\mathrm{ZF}^\mathrm{KT}(\Delta,t) = \frac{1}{3} + \frac{2}{3}(1-\Delta^2t^2)\exp\left(-{\textstyle\frac{1}{2}}\Delta^2t^2\right)
\end{equation}
is the ZF Kubo-Toyabe (KT) form expected~\cite{KUBO67,*HAYANO79} from an isotropic Gaussian distribution of randomly oriented static or quasistatic local fields at muon sites. In Eqs.~(\ref{eq:ZF-asym}) and (\ref{eq:KT}) $\Delta/\gamma_{\mu}$ is the the rms width of this distribution ($\gamma_{\mu} = 2\pi \times 135.53$~MHz/T is the muon gyromagnetic ratio), and $\lambda_\mathrm{ZF}$ in Eq.~(\ref{eq:ZF-asym}) is the rate of exponential damping due to dynamic fluctuations of the local muon fields.

At 1.2~K the ZF-$\mu$SR spectra (Fig.~\ref{fig1}) are nearly of the KT form; the exponential damping is weak. This indicates that the relaxation is dominated by muon precession in quasistatic nuclear dipolar fields. Fits of Eq.~(\ref{eq:ZF-asym}) to the 1.2-K data yield $\Delta =0.32(1)~\mu\mathrm{s}^{-1}$ and $0.27(2)~\mu$s$^{-1}$ for $\mathbf{P}_\mu\,{\parallel}\,\mathbf{b}$ and $\mathbf{c}$, respectively. Implanted muons are expected to be immobile at temperatures below $\sim$100~K, in which case $\Delta$ is independent of temperature. At low temperatures the relaxation is considerably faster and more nearly exponential than at 1.2~K (Fig.~\ref{fig1}): $\lambda_\mathrm{ZF}$ has increased with decreasing temperature, whereas $\Delta$ is essentially unchanged (data not shown).

\begin{figure}[t]
 \includegraphics[clip=,width=0.45\textwidth]{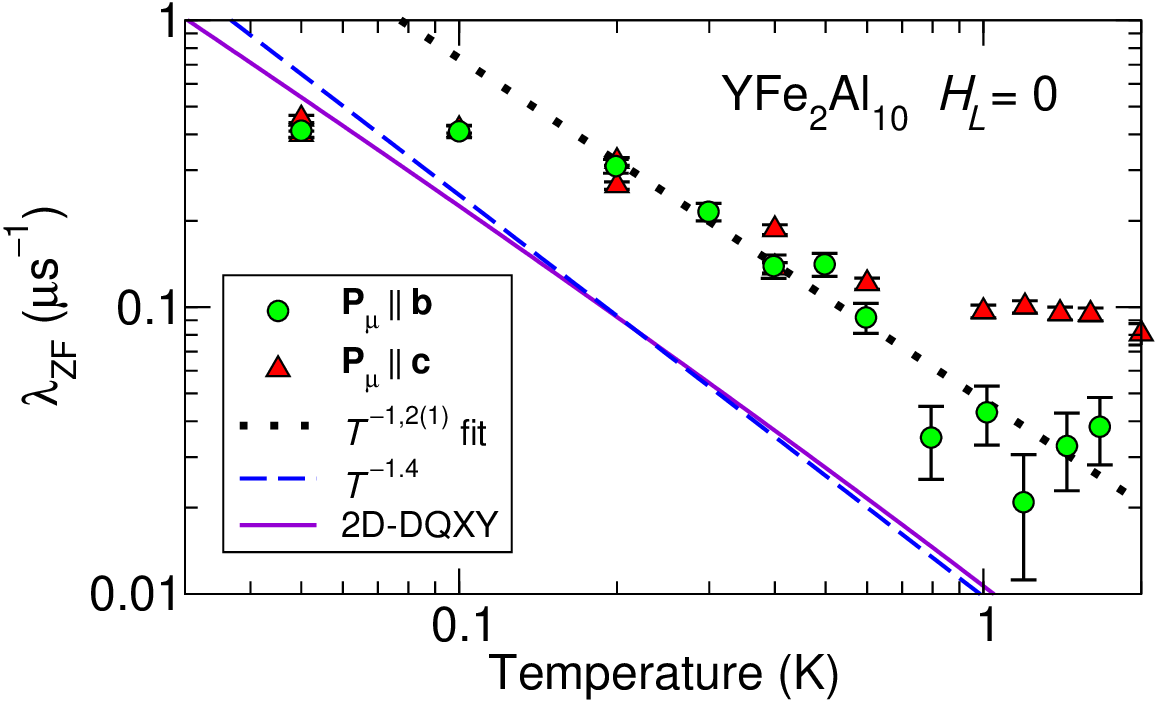}
 \caption{Temperature dependence of the zero-field muon spin relaxation rate~$\lambda_\mathrm{ZF}$. Circles: $\mathbf{P}_\mu\,{\parallel}\,\mathbf{b}$. Triangles: $\mathbf{P}_\mu\,{\parallel}\,\mathbf{c}$. Dotted line: power-law fit to the data for $\mathbf{P}_\mu\,{\parallel}\,\mathbf{b}$, $0.1~\text{K} < T \leqslant 1$~K. Dashed line: $T^{-1.4}$ power law. Solid curve: 2D-DQXY functional form~\cite{VARMA17u}. The latter two curves are normalized to neutron scattering data (see text).}
 \label{fig2}
\end{figure}

Figure~\ref{fig2} shows the temperature dependence of $\lambda_\mathrm{ZF}$.
From ${\sim}0.05$--$0.1~\mu\mathrm{s}^{-1}$ above 1~K, $\lambda_\mathrm{ZF}$ increases by an order of magnitude with decreasing temperature and then saturates below $\sim$0.1~K\@. Below $\sim$0.5~K the relaxation is isotropic. Anisotropy develops at higher temperatures, where the relaxation is dominated by a mechanism or mechanisms other than quantum criticality.

The temperature dependence of the bulk magnetic susceptibility with field in the $ac$ plane exhibits a $T^{-1.4}$ power-law divergence~\cite{WU14}. The dotted line in Fig.~\ref{fig2} is a power-law fit to the data for $\mathbf{P}_\mu\,{\parallel}\,\mathbf{b}$, $0.1~\text{K} < T \leqslant 1$~K; this yields an exponent~$-1.2(1)$, not far from the susceptibility value. The dashed line in Fig.~\ref{fig2} is from a $T^{-1.4}$ power-law scaling scenario~\cite{GANNON17u}, and the solid curve is from from the 2D-DQXY model~\cite{VARMA17u}. The magnitudes of both predictions are normalized using neutron scattering data, as discussed in detail in Sec.~\ref{sec:disc}. In this temperature range the predictions are nearly the same, and (although the absolute magnitudes are a factor 3--4 too small) quite comparable to the observed data, particularly for $\mathbf{P}_\mu\,{\parallel}\,\mathbf{b}$.

In a number of quantum ferromagnetic materials the temperature dependence of the muon spin relaxation rate also obeys a power law (above the Curie temperature if there is ferromagnetic ordering). Experimental values of the exponent are $-0.8$ in CePd$_{0.15}$Rh$_{0.85}$~\cite{ADROJA08}, $-0.33$ in YbCu$_{4.4}$Au$_{0.6}$ \cite{CARRETTA09}, $-0.4$ to $-0.5$ in YNi$_4$P$_2$~\cite{SPEHLING12}, and $-0.01$ to $-0.13$ in YbNi$_4$(P$_{1−x}$As$_x$)$_2$~\cite{SARKAR17}. In YFe$_2$Al$_{10}$ the exponent magnitude is considerably larger than in these materials.

\subsection{\boldmath Longitudinal-field $\mu$SR}

\begin{figure}[t]
 \includegraphics[clip=,width=0.45\textwidth]{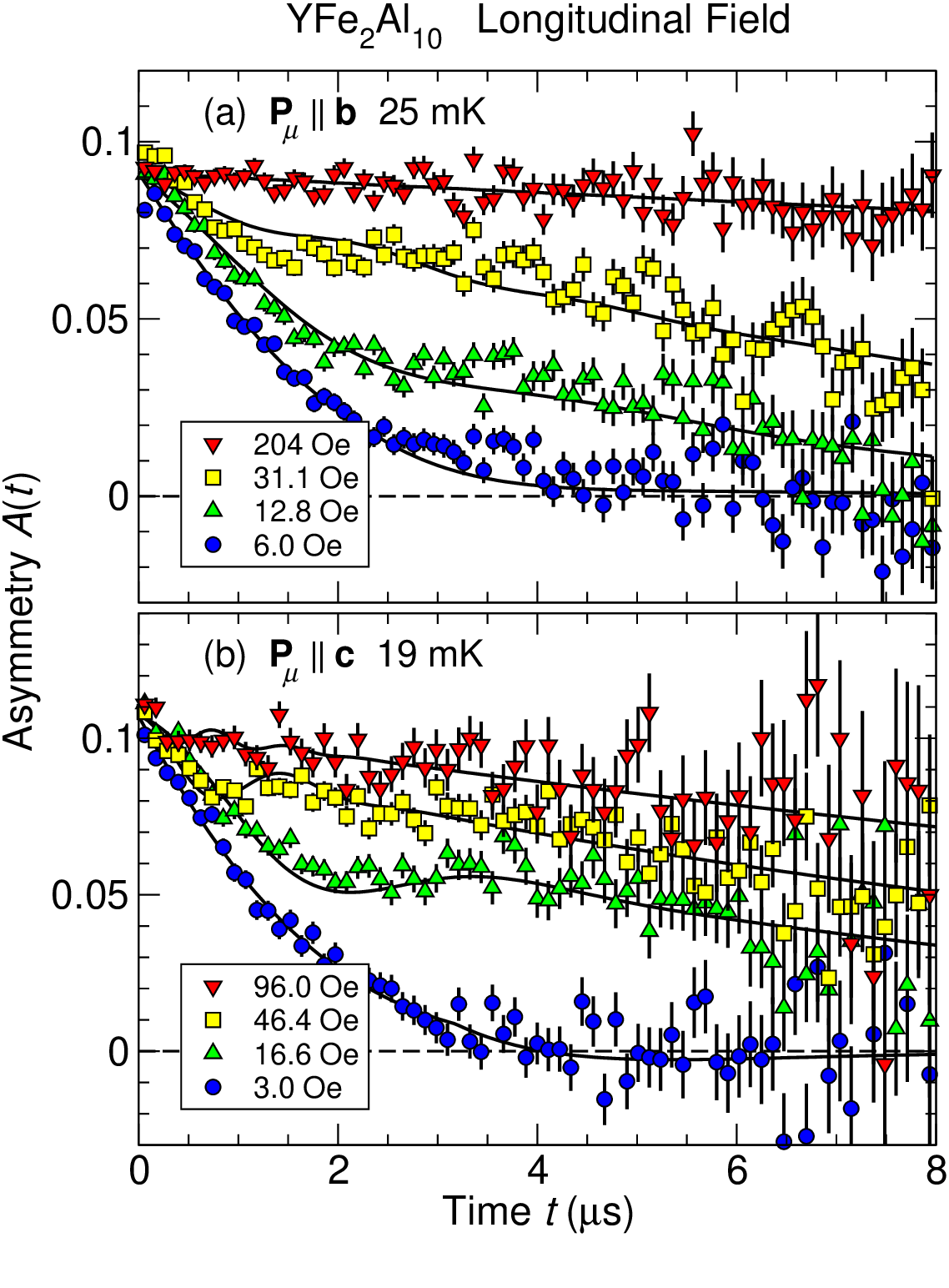}
 \caption{Dependence of muon asymmetry relaxation function $A(t)$ on longitudinal field~$H_L$ in oriented single crystals of YFe$_2$Al$_{10}$. (a)~$\mathbf{H}_L\,{\parallel}\,\mathbf{b}$. (b)~$\mathbf{H}_L\,{\parallel}\, \mathbf{c}$. Curves: fits of exponentially-damped Gaussian static Kubo-Toyabe functions to the data (see text). Decoupling of static relaxation and slowing of dynamic relaxation both contribute to the field dependence.}
 \label{fig3}
\end{figure}

LF-$\mu$SR relaxation rates were measured in YFe$_{2}$Al$_{10}$ for applied magnetic fields up to 200~Oe at a number of temperatures. Figures~\ref{fig3}(a) and \ref{fig3}(b) show LF-$\mu$SR asymmetry spectra at base temperatures for $\mathbf{H}_L\,{\parallel}\,\mathbf{b}$ and $\mathbf{H}_L\,{\parallel}\,\mathbf{c}$, respectively.
The LF-$\mu$SR spectra are well described by an exponentially damped static relaxation function similar to that of Eq.~(\ref{eq:ZF-asym}), except that the static Gaussian KT function~$G_\mathrm{LF}^\mathrm{KT}(\Delta,t)$ for a longitudinal magnetic field~\cite{HAYANO79} is used. The majority of the field dependence seen in Fig.~\ref{fig3} is due to ``decoupling'' of the static relaxation by the field~\cite{HAYANO79}, but the dynamic relaxation also slows with increasing field.

\begin{figure}[t]
\begin{center}
 \includegraphics[clip=,width=0.45\textwidth]{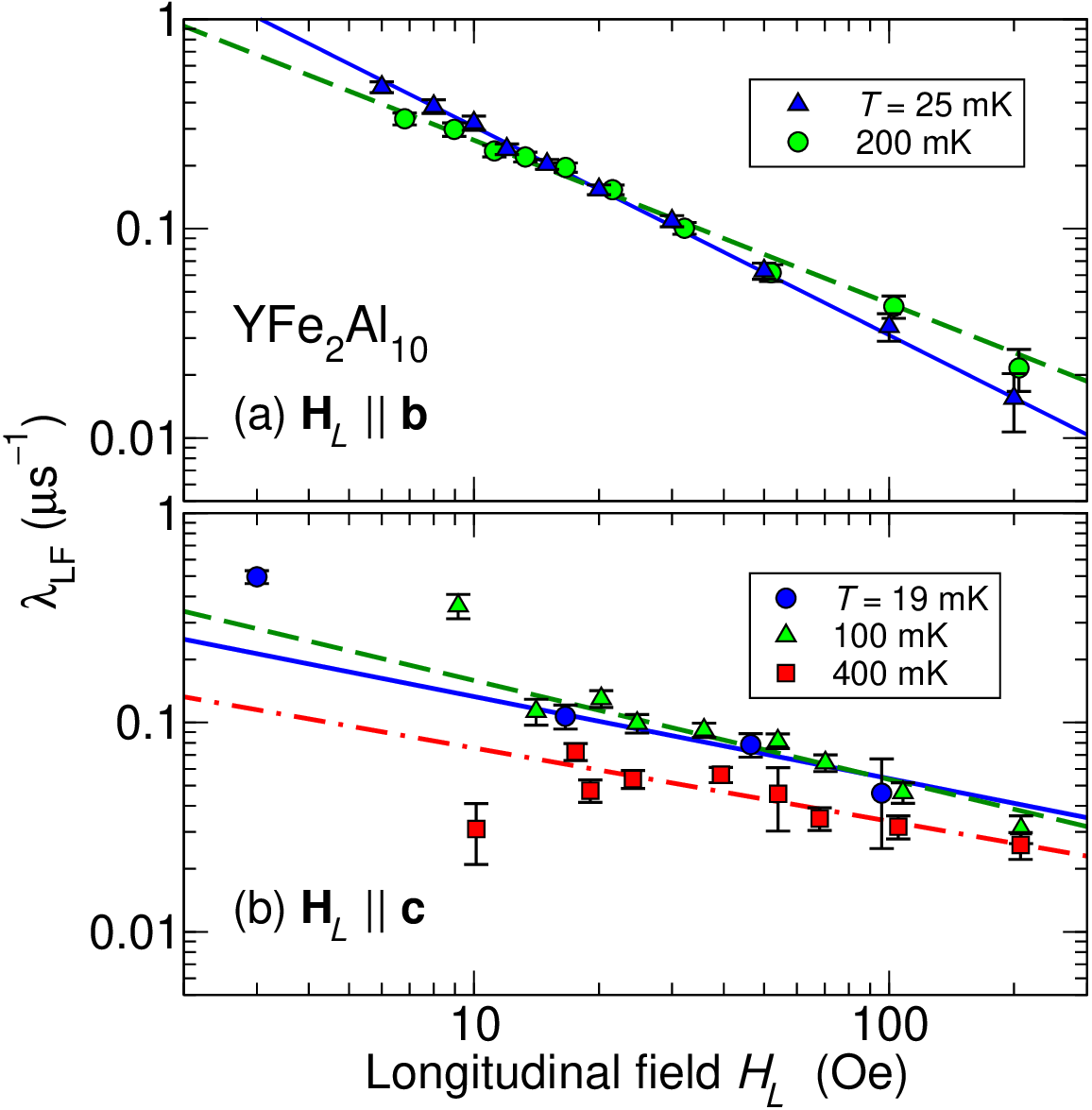}
 \caption{Field dependence of the LF dynamic relaxation rate~$\lambda_\mathrm{LF}(H)$ at low temperatures. Lines are power-law fits to the data. (a)~longitudinal field~$\mathbf{H}_L$ parallel to crystal axis $\mathbf{b}$. Solid line: $T = 25$~mK, $\text{slope} = -1.00(5)$. Dashed line: $T = 200$~mK, $\text{slope} = -0.78(4)$. (b)~$\mathbf{H}_L\,\parallel\,\mathbf{c}$. Lines are fit to data above 10~Oe. Solid line: $T = 19$~mK, $\text{slope} = -0.39(11)$. Dashed line: $T = 100$~mK, $\text{slope} = -0.47(7)$. Dash-dot line: $T = 400$~mK, $\text{slope} = -0.35(8)$.}
 \label{fig4}
\end{center}
\end{figure}

Figure~\ref{fig4} shows the field dependence of $\lambda_\mathrm{LF}$ from LF-$\mu$SR data for longitudinal field~$\mathbf{H}_L$ oriented along the $b$ and $c$ crystal axes.
The lines are power-law fits to all ($\mathbf{H}_L\,{\parallel}\,\mathbf{b}$) or part ($\mathbf{H}_L\,{\parallel}\,\mathbf{c}$) of the data. There is a general tendency for the power-law exponent to decrease with increasing temperature. For $\mathbf{H}_L\,{\parallel}\,\mathbf{c}$ the data deviate from power laws at low fields, and the change of $\lambda_\mathrm{LF}$ is small enough to cast doubt on the uniqueness of power-law fits. The behavior of $\lambda_\mathrm{LF}$ is clearly anisotropic. This is not well understood, but may be due to the anisotropy in the Fe$^{2+}$ dipolar fields at muon sites~\footnote{In zero applied field the muon Zeeman splitting is due to nuclear dipolar fields, which are randomly oriented and tend to average out any anisotropy.}.

The field dependence of $\lambda_\mathrm{LF}$ might suggest that it originates from a Lorentzian contribution to the static field distribution and is thus decoupled by the field, as in PrPt$_4$Ge$_{12}$~\cite{MAISURADZE10}. This seems unlikely, however. Particularly for $\mathbf{H}_L\,{\parallel}\,\mathbf{b}$, the field dependence of $\lambda$ is accurately represented by a power law for fields above and below ${\sim}5\lambda_\mathrm{ZF} \approx 20$~Oe [Fig.~\ref{fig4}(a)], where decoupling is complete~\cite{HAYANO79}. This would not be expected if the mechanism for $\lambda$ were different in low and high fields.

\section{DISCUSSION} \label{sec:disc}

The muon spin relaxation rate is related to the imaginary component~$\chi''(\mathbf{q},\omega_\mu,T)$ of the dynamic susceptibility, as a consequence of the fluctuation-dissipation theorem~\cite{MORIYA63}:
\begin{equation} \label{eq:general}
\lambda(\omega_\mu,T) = \frac{2\hbar\gamma_\mu^2}{g^2\mu_B^2}\left(\frac{k_BT}{\hbar\omega_\mu}\right)\sum_\mathbf{q} A_\mathbf{q} A_{-\mathbf{q}} \chi''(\mathbf{q},\omega_\mu,T) \,,
\end{equation}
where $\omega_\mu = \gamma_\mu H_L$ is the muon Zeeman frequency and $A_\mathbf{q}$ is the spatial Fourier transform of the coupling magnetic field between electronic fluctuations and probe spins. Furthermore, the dynamic structure factor~$S(\omega,T)$ is given by~\cite{GANNON17u}
\begin{equation} \label{eq:Schi}
S(\mathbf{q},\omega,T) = \frac{2(n + 1)}{\pi g^2\mu_B^2}\chi''(\mathbf{q},\omega,T) \,,
\end{equation}
where $n + 1 = [1 - \exp(-\hbar\omega/k_BT)]^{-1}$ is the Bose or detailed balance factor. In the limit $\omega \ll T$, $n+1 \to k_BT/\hbar\omega$, which is the factor in parentheses in Eq.~(\ref{eq:general}). Thus $S$ and $\lambda$ are closely related.

If the electronic spin fluctuations are not spatially correlated, as is the case in YFe$_2$Al$_{10}$~\cite{GANNON17u}, then $\chi''(\mathbf{q},\omega_\mu,T)$ is independent of $\mathbf{q}$, and from Eq.~(\ref{eq:general})
\begin{equation} \label{eq:uncorr}
\lambda(\omega_\mu,T) = \frac{2\hbar\gamma_\mu^2|A|^2}{g^2\mu_B^2}\left(\frac{k_BT}{\hbar\omega_\mu}\right)\chi''(\omega_\mu,T)\,,
\end{equation}
where $|A^2| = \sum_\mathbf{q} A_\mathbf{q} A_{-\mathbf{q}}$. Then
\begin{equation} \label{eq:Slambda}
S(\omega_\mu,T) = \lambda(\omega_\mu,T)/\pi\hbar\gamma_\mu^2|A|^2 \,,
\end{equation}
and the field dependence of $\lambda(H_L,T)$ directly probes the frequency dependence of $S(\omega,T)$ at the low muon frequencies. In effect we are sweeping the muon Zeeman frequency through the noise power spectrum of the fluctuations~\cite{KEREN96,KEREN01}.

For this analysis to be valid, the applied field must be low enough that the fluctuating spin dynamics are not affected. This appears to be the case in YFe$_2$Al$_{10}$~\cite{WU14,GANNON17u}, and we attribute the field dependence of $\lambda_\mathrm{LF}$ to the frequency dependence of the dynamic structure factor.

Thus we can make a quantitative comparison between neutron and $\mu$SR results, notwithstanding the large difference in frequencies probed in the two experiments. In neutron scattering $\hbar\omega/k_B$ is typically of the order of kelvin or greater, whereas $\hbar\gamma_\mu/k_B \sim 10^{-6}$~K/Oe.

We consider two models for $\chi''(\omega,T)$. Motivated by magnetic susceptibility results and scaling arguments from earlier studies~\cite{WU14}, Gannon \textit{et al.}~\cite{GANNON17u} have shown that their data can be fit by either a temperature-independent $\omega^{-1.4}$ power law or the $\omega/T$ scaling form (in units where $\hbar/k_B = 1$)
\begin{equation} \label{eq:pwrlaw}
\chi''(\omega,T) \propto  [\omega^2 + (\pi T)^2]^{-1.4/2} \tanh(\omega/T) \,.
\end{equation}
Alternatively, a treatment of critical spin dynamics within the 2D-DQXY model~\cite{VARMA17u} suggests the form
\begin{equation} \label{eq:qddxy}
\chi''(\omega,T) \propto  \frac{\log^2\left\{[\omega^2 + (\pi T)^2]^{1/2}/\omega_c\right\}}{[\omega^2 + (\pi T)^2]^{1/2}}\tanh(\omega/T) \,,
\end{equation}
where $\omega_c$ is a high-frequency cutoff. The combination of temperature and frequency avoids a divergence in the Kramers-Kronig relation between $\chi''$ and $\chi'$~\cite{VARMA17u,GANNON17u} that would occur without the temperature cutoff in the frequency dependence. Over the temperature range of the experiments the frequency and temperature dependencies of Eqs.~(\ref{eq:pwrlaw}) and (\ref{eq:qddxy}) are essentially indistinguishable at $\mu$SR frequencies, as is also the case at meV neutron scattering energies as noted above.

To obtain quantitative values of $S$ from relaxation rates we must determine $|A|^2$. We have calculated dipolar fields due to Fe moments in the $ac$ plane at two candidate muon spin sites~\cite{[{This calculation uses the $0,0,0$ and $0,\frac{1}{2},\frac{1}{4}$ muon sites proposed for iso\-structural Ce(Rh,Ru)$_2$Al$_{10}$, cf.\ }] ADAM14} from lattice sums in the orthorhombic YbFe$_2$Al$_{10}$ structure (space group~\textit{Cmcm}, no.~63). The rms dipolar coupling field~$A_\mathrm{rms} = |A^2|^{1/2}$ varies between 400 and 800 G/$\mu_B$, depending on the site and the muon spin direction. Then $\gamma_\mu A_\mathrm{rms} \approx 5 \times 10^7~\mathrm{s}^{-1}/\mu_B$ to within a factor of two. As an example, the measured value of $\lambda_\mathrm{ZF}$ at 25~mK is about $4 \times 10^5\ \mathrm{s}^{-1}$. With $\hbar = 6.582 \times 10^{-13}$~meV~s, the effective structure factor from Eq.~(\ref{eq:Slambda}) is
\begin{equation}
S \approx \frac{4 \times 10^5}{\pi\hbar(5 \times 10^7)^2} \approx 80\ \mu_B^2/\text{meV-Fe}
\end{equation}
to within a factor of four.

We next compare these model results with our data.

\paragraph{Zero-field $\mu$SR.} As noted in Sec.~\ref{par:ZF}, the observed temperature dependence of $\lambda_\mathrm{ZF}(T)$ is compatible with either of the above scenarios. The absolute value of $\lambda_ZF$ is underestimated by a factor of 3--4 (Fig.~\ref{fig2}) but this may be considered as semi-quantitative agreement, considering the accumulated uncertainty involved in determining the normalization factor and the effect of noncritical relaxation mechanisms.

\paragraph{Longitudinal-field $\mu$SR.} An order-of-magnitude suppression of $\lambda_\mathrm{LF}$ is observed in fields as low as 50~Oe (Fig.~\ref{fig4}). A possible scenario for this might involve the preasymptotic/asymptotic crossover predicted for disordered quantum ferromagnets~\cite{BRANDO16,KIRKPATRICK14}. But Park \textit{et al.}~\cite{PARK11} have argued that YFe$_2$Al$_{10}$ is a clean system, for which the theory of Ref.~\cite{KIRKPATRICK14} would not be appropriate. Furthermore, field dependence of the Fe$^{2+}$ spin fluctuation spectrum seems unlikely, since both the uniform~\cite{WU14} and dynamic~\cite{GANNON17u} spin susceptibilities are unaffected by fields of this magnitude. Instead, as discussed above, we attribute the field dependence of $\lambda_\mathrm{LF}$ to the frequency dependence of $\chi''$ and hence of $S$.

\begin{figure}[t]
\vspace{10pt}
\begin{center}
 \includegraphics[clip=,width=0.45\textwidth]{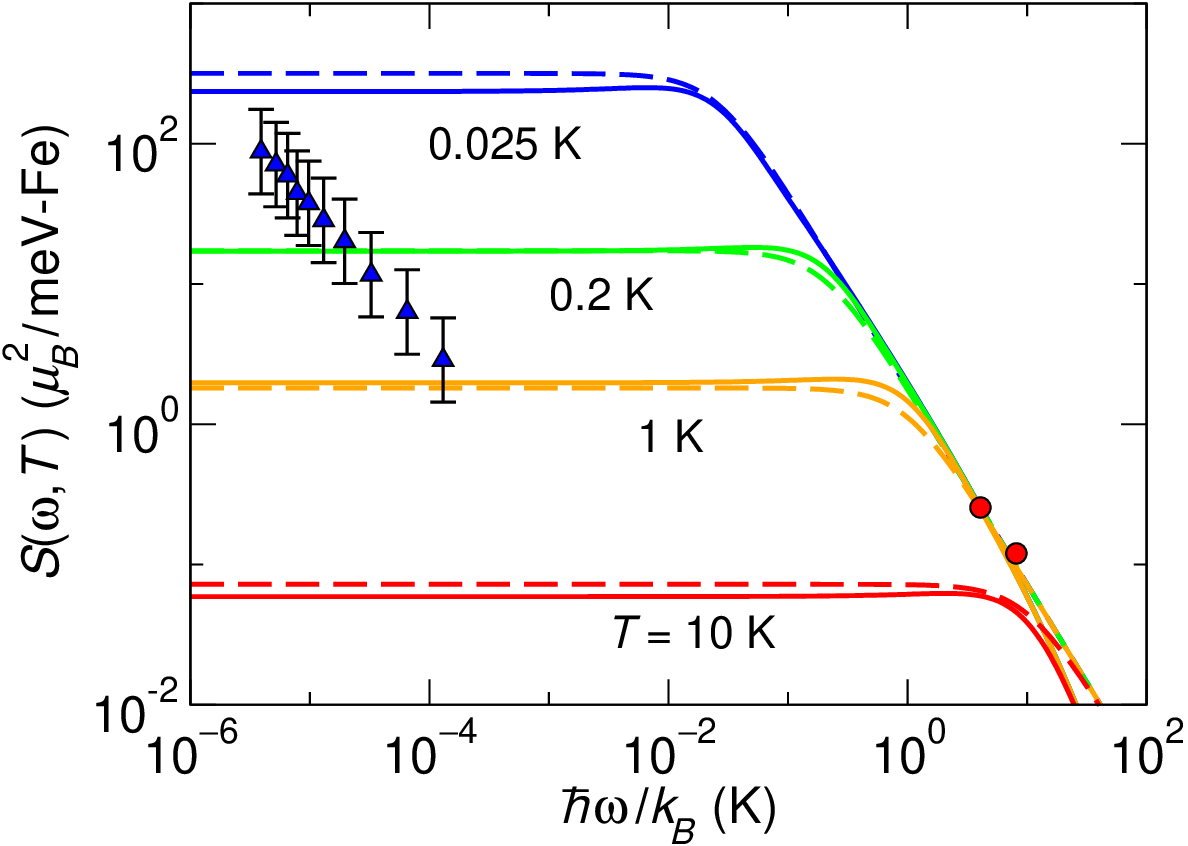}
 \caption{Frequency dependence of the dynamic structure factor~$S(\omega,T)$. Circles: $S(\omega)$ from neutron scattering~\cite{GANNON17u}. Triangles: $S(\omega)$ from $\mu$SR relaxation rates for $\mathbf{H}_L\,{\parallel}\,\mathbf{b}$ and $T = 25$~mK [Fig.~\ref{fig4}(a) and Eq.~(\ref{eq:Slambda})]. Solid curves: $S(\omega,T)$ from 2D-DQXY model~\cite{VARMA17u} \{Eq.~(\ref{eq:qddxy}), $\omega_c = 161$~K~\cite{GANNON17u}\}. Dashed curves: scaling power law [Eq.~(\ref{eq:pwrlaw})]. Both sets of curves are normalized to the neutron scattering data.}
 \label{fig5}
\end{center}
\end{figure}

Figure~\ref{fig5} gives the calculated frequency dependence of $S(\omega,T)$ from the scaling scenarios at various temperatures, together with results from neutron scattering~\cite{GANNON17u} and from LF-$\mu$SR relaxation data and Eq.~(\ref{eq:Slambda})
(the error bars are due to the uncertainty in the estimate of the rms coupling field noted above, which is much greater than the statistical error in $\lambda_\mathrm{LF}$). The curves are from Eqs.~(\ref{eq:pwrlaw}) and (\ref{eq:qddxy}) ($\omega_c = 161$~K~\cite{VARMA17u}), normalized to the neutron scattering data and evaluated at various temperatures. For $\omega \ll T$ the scaling scenarios predict no frequency dependence of $S$ and hence no field dependence of $\lambda_\mathrm{LF}$ for low fields, contrary to the considerable observed power-law field dependence.

This discrepancy is not understood at present. We note that the data seem to suggest a product form~$S(\omega,T) \propto \omega^{-\Gamma} T^{-\Delta}$ at ultralow frequencies, rather than the $\omega^2 + (\pi T)^2$ dependence of the power-law and 2D-DQXY model results. Such a product preserves $\omega/T$ scaling, but the resultant divergence in $\chi''(\omega)$ leads to the problem with the Kramers-Kronig relation discussed above.

It should perhaps also be noted that the suppression of $\lambda_\mathrm{LF}$ with field at low temperatures cannot be due to mechanisms other than that which increases $\lambda_\mathrm{ZF}$ at low temperatures. Such mechanisms would increase the rate rather than decreasing it.

\section{SUMMARY}

Dynamic muon spin relaxation has been studied in the quasi-2D nearly-ferromagnetic compound~YFe$_2$Al$_{10}$. The relaxation behavior indicates that there is no static electronic magnetism, ordered or disordered, down to 19~mK\@. In zero applied field the dynamic muon spin relaxation rate~$\lambda_\mathrm{ZF}$ is strongly enhanced below 1~K, saturating at ${\sim}0.4~\mu\mathrm{s}^{-1}$ below 0.1~K\@. In the temperature range 0.1--1.0~K a power law $\lambda_\mathrm{ZF}(T) \propto T^{-1.2(1)}$ was observed. This power and the magnitude of $\lambda_\mathrm{ZF}$ are in semi-quantitative agreement with extrapolations of power-law scaling or the 2D-DQXY model for 2D ferromagnetic quantum critical fluctuations. At 25~mK $\lambda_\mathrm{LF}(\mathbf{H}_L\,{\parallel}\,\mathbf{b})$ exhibits a power-law dependence on field, with exponent~$-1.00(5)$. This is in extreme disagreement with the frequency independence expected from power-law scaling or the 2D-DQXY model. We conclude that neither of these fully captures the low-frequency spin dynamics associated with the QCP in YFe$_2$Al$_{10}$, and that more work is necessary to understand this elusive system.

\begin{acknowledgments}

We are grateful to the $\mu$SR support staff at TRIUMF and PSI for their help during the experiments, and to C.~M. Varma for many hours of useful discussion. The research performed in this study was supported by the National Key Research and Development Program of China (Nos.~2017YFA0303104 and 2016YFA0300503), and the National Natural Science Foundation of China under Grant nos.~11474060 and 11774061. Work at Texas A$\&$M University (M. C. A) and Stony Brook University (M. C. A. and L. S. W) was supported by the National Science Foundation DMR-1310008. Work at CSULA was funded by the U.S. NSF DMR/PREM-1523588. Research at CSU Fresno was supported by NSF DMR-1506677. Research at UC Riverside (UCR) was supported by the UCR Academic Senate.

\end{acknowledgments}
%


\begin{thebibliography}{35}%
\makeatletter
\providecommand \@ifxundefined [1]{%
 \@ifx{#1\undefined}
}%
\providecommand \@ifnum [1]{%
 \ifnum #1\expandafter \@firstoftwo
 \else \expandafter \@secondoftwo
 \fi
}%
\providecommand \@ifx [1]{%
 \ifx #1\expandafter \@firstoftwo
 \else \expandafter \@secondoftwo
 \fi
}%
\providecommand \natexlab [1]{#1}%
\providecommand \enquote  [1]{``#1''}%
\providecommand \bibnamefont  [1]{#1}%
\providecommand \bibfnamefont [1]{#1}%
\providecommand \citenamefont [1]{#1}%
\providecommand \href@noop [0]{\@secondoftwo}%
\providecommand \href [0]{\begingroup \@sanitize@url \@href}%
\providecommand \@href[1]{\@@startlink{#1}\@@href}%
\providecommand \@@href[1]{\endgroup#1\@@endlink}%
\providecommand \@sanitize@url [0]{\catcode `\\12\catcode `\$12\catcode
  `\&12\catcode `\#12\catcode `\^12\catcode `\_12\catcode `\%12\relax}%
\providecommand \@@startlink[1]{}%
\providecommand \@@endlink[0]{}%
\providecommand \url  [0]{\begingroup\@sanitize@url \@url }%
\providecommand \@url [1]{\endgroup\@href {#1}{\urlprefix }}%
\providecommand \urlprefix  [0]{URL }%
\providecommand \Eprint [0]{\href }%
\providecommand \doibase [0]{http://dx.doi.org/}%
\providecommand \selectlanguage [0]{\@gobble}%
\providecommand \bibinfo  [0]{\@secondoftwo}%
\providecommand \bibfield  [0]{\@secondoftwo}%
\providecommand \translation [1]{[#1]}%
\providecommand \BibitemOpen [0]{}%
\providecommand \bibitemStop [0]{}%
\providecommand \bibitemNoStop [0]{.\EOS\space}%
\providecommand \EOS [0]{\spacefactor3000\relax}%
\providecommand \BibitemShut  [1]{\csname bibitem#1\endcsname}%
\let\auto@bib@innerbib\@empty
\bibitem [{\citenamefont {Brando}\ \emph {et~al.}(2016)\citenamefont {Brando},
  \citenamefont {Belitz}, \citenamefont {Grosche},\ and\ \citenamefont
  {Kirkpatrick}}]{BRANDO16}%
  \BibitemOpen
  \bibfield  {author} {\bibinfo {author} {\bibfnamefont {M.}~\bibnamefont
  {Brando}}, \bibinfo {author} {\bibfnamefont {D.}~\bibnamefont {Belitz}},
  \bibinfo {author} {\bibfnamefont {F.~M.}\ \bibnamefont {Grosche}}, \ and\
  \bibinfo {author} {\bibfnamefont {T.~R.}\ \bibnamefont {Kirkpatrick}},\
  }\href {\doibase 10.1103/RevModPhys.88.025006} {\bibfield  {journal}
  {\bibinfo  {journal} {Rev. Mod. Phys.}\ }\textbf {\bibinfo {volume} {88}},\
  \bibinfo {pages} {025006} (\bibinfo {year} {2016})}\BibitemShut {NoStop}%
\bibitem [{\citenamefont {Belitz}\ \emph {et~al.}(2005)\citenamefont {Belitz},
  \citenamefont {Kirkpatrick},\ and\ \citenamefont {Vojta}}]{BKV05}%
  \BibitemOpen
  \bibfield  {author} {\bibinfo {author} {\bibfnamefont {D.}~\bibnamefont
  {Belitz}}, \bibinfo {author} {\bibfnamefont {T.~R.}\ \bibnamefont
  {Kirkpatrick}}, \ and\ \bibinfo {author} {\bibfnamefont {T.}~\bibnamefont
  {Vojta}},\ }\href {\doibase 10.1103/RevModPhys.77.579} {\bibfield  {journal}
  {\bibinfo  {journal} {Rev. Mod. Phys.}\ }\textbf {\bibinfo {volume} {77}},\
  \bibinfo {pages} {579} (\bibinfo {year} {2005})}\BibitemShut {NoStop}%
\bibitem [{\citenamefont {v.~L{\"o}hneysen}\ \emph {et~al.}(2007)\citenamefont
  {v.~L{\"o}hneysen}, \citenamefont {Rosch}, \citenamefont {Vojta},\ and\
  \citenamefont {W{\"o}lfle}}]{LOHNEYSEN07}%
  \BibitemOpen
  \bibfield  {author} {\bibinfo {author} {\bibfnamefont {H.}~\bibnamefont
  {v.~L{\"o}hneysen}}, \bibinfo {author} {\bibfnamefont {A.}~\bibnamefont
  {Rosch}}, \bibinfo {author} {\bibfnamefont {M.}~\bibnamefont {Vojta}}, \ and\
  \bibinfo {author} {\bibfnamefont {P.}~\bibnamefont {W{\"o}lfle}},\ }\href
  {\doibase 10.1103/RevModPhys.79.1015} {\bibfield  {journal} {\bibinfo
  {journal} {Rev. Mod. Phys.}\ }\textbf {\bibinfo {volume} {79}},\ \bibinfo
  {pages} {1015} (\bibinfo {year} {2007})}\BibitemShut {NoStop}%
\bibitem [{\citenamefont {Wu}\ \emph {et~al.}(2014)\citenamefont {Wu},
  \citenamefont {Kim}, \citenamefont {Park}, \citenamefont {Tsvelik},\ and\
  \citenamefont {Aronson}}]{WU14}%
  \BibitemOpen
  \bibfield  {author} {\bibinfo {author} {\bibfnamefont {L.~S.}\ \bibnamefont
  {Wu}}, \bibinfo {author} {\bibfnamefont {M.~S.}\ \bibnamefont {Kim}},
  \bibinfo {author} {\bibfnamefont {K.}~\bibnamefont {Park}}, \bibinfo {author}
  {\bibfnamefont {A.~M.}\ \bibnamefont {Tsvelik}}, \ and\ \bibinfo {author}
  {\bibfnamefont {M.~C.}\ \bibnamefont {Aronson}},\ }\href {\doibase
  10.1103/PhysRevB.87.064502} {\bibfield  {journal} {\bibinfo  {journal} {Proc.
  Nat. Acad. Sci.}\ }\textbf {\bibinfo {volume} {111}},\ \bibinfo {pages}
  {14088} (\bibinfo {year} {2014})}\BibitemShut {NoStop}%
\bibitem [{\citenamefont {M.~T.~Thiede}\ \emph {et~al.}(1998)\citenamefont
  {M.~T.~Thiede}, \citenamefont {Ebel},\ and\ \citenamefont
  {Jeitschko}}]{TEJ98}%
  \BibitemOpen
  \bibfield  {author} {\bibinfo {author} {\bibfnamefont {V.}~\bibnamefont
  {M.~T.~Thiede}}, \bibinfo {author} {\bibfnamefont {T.}~\bibnamefont {Ebel}},
  \ and\ \bibinfo {author} {\bibfnamefont {W.}~\bibnamefont {Jeitschko}},\
  }\href {\doibase 10.1039/A705854C} {\bibfield  {journal} {\bibinfo  {journal}
  {J. Mater. Chem.}\ }\textbf {\bibinfo {volume} {8}},\ \bibinfo {pages} {125}
  (\bibinfo {year} {1998})}\BibitemShut {NoStop}%
\bibitem [{\citenamefont {Strydom}\ and\ \citenamefont
  {Peratheepan}(2010)}]{STRYDOM10}%
  \BibitemOpen
  \bibfield  {author} {\bibinfo {author} {\bibfnamefont {A.~M.}\ \bibnamefont
  {Strydom}}\ and\ \bibinfo {author} {\bibfnamefont {P.}~\bibnamefont
  {Peratheepan}},\ }\href {\doibase 10.1002/pssr.201004365} {\bibfield
  {journal} {\bibinfo  {journal} {Phys. Status Solidi RRL}\ }\textbf {\bibinfo
  {volume} {4}},\ \bibinfo {pages} {356} (\bibinfo {year} {2010})}\BibitemShut
  {NoStop}%
\bibitem [{\citenamefont {Strydom}\ \emph {et~al.}(2011)\citenamefont
  {Strydom}, \citenamefont {Peratheepan}, \citenamefont {Sarkar}, \citenamefont
  {Baenitz},\ and\ \citenamefont {Steglich}}]{STRYDOM11}%
  \BibitemOpen
  \bibfield  {author} {\bibinfo {author} {\bibfnamefont {A.~M.}\ \bibnamefont
  {Strydom}}, \bibinfo {author} {\bibfnamefont {P.}~\bibnamefont
  {Peratheepan}}, \bibinfo {author} {\bibfnamefont {R.}~\bibnamefont {Sarkar}},
  \bibinfo {author} {\bibfnamefont {M.}~\bibnamefont {Baenitz}}, \ and\
  \bibinfo {author} {\bibfnamefont {F.}~\bibnamefont {Steglich}},\ }\href@noop
  {} {\bibfield  {journal} {\bibinfo  {journal} {J. Phys. Soc. Japan}\ }\textbf
  {\bibinfo {volume} {80}},\ \bibinfo {pages} {1} (\bibinfo {year}
  {2011})}\BibitemShut {NoStop}%
\bibitem [{\citenamefont {Park}\ \emph {et~al.}(2011)\citenamefont {Park},
  \citenamefont {Wu}, \citenamefont {Janssen}, \citenamefont {Kim},
  \citenamefont {Marques},\ and\ \citenamefont {Aronson}}]{PARK11}%
  \BibitemOpen
  \bibfield  {author} {\bibinfo {author} {\bibfnamefont {K.}~\bibnamefont
  {Park}}, \bibinfo {author} {\bibfnamefont {L.~S.}\ \bibnamefont {Wu}},
  \bibinfo {author} {\bibfnamefont {Y.}~\bibnamefont {Janssen}}, \bibinfo
  {author} {\bibfnamefont {M.~S.}\ \bibnamefont {Kim}}, \bibinfo {author}
  {\bibfnamefont {C.}~\bibnamefont {Marques}}, \ and\ \bibinfo {author}
  {\bibfnamefont {M.~C.}\ \bibnamefont {Aronson}},\ }\href {\doibase
  10.1103/PhysRevB.84.094425} {\bibfield  {journal} {\bibinfo  {journal} {Phys.
  Rev. B}\ }\textbf {\bibinfo {volume} {84}},\ \bibinfo {pages} {094425}
  (\bibinfo {year} {2011})}\BibitemShut {NoStop}%
\bibitem [{\citenamefont {Khuntia}\ \emph {et~al.}(2012)\citenamefont
  {Khuntia}, \citenamefont {Strydom}, \citenamefont {Wu}, \citenamefont
  {Aronson}, \citenamefont {Steglich},\ and\ \citenamefont
  {Baenitz}}]{KHUNTIA12}%
  \BibitemOpen
  \bibfield  {author} {\bibinfo {author} {\bibfnamefont {P.}~\bibnamefont
  {Khuntia}}, \bibinfo {author} {\bibfnamefont {A.~M.}\ \bibnamefont
  {Strydom}}, \bibinfo {author} {\bibfnamefont {L.~S.}\ \bibnamefont {Wu}},
  \bibinfo {author} {\bibfnamefont {M.~C.}\ \bibnamefont {Aronson}}, \bibinfo
  {author} {\bibfnamefont {F.}~\bibnamefont {Steglich}}, \ and\ \bibinfo
  {author} {\bibfnamefont {M.}~\bibnamefont {Baenitz}},\ }\href {\doibase
  10.1103/PhysRevB.86.220401} {\bibfield  {journal} {\bibinfo  {journal} {Phys.
  Rev. B}\ }\textbf {\bibinfo {volume} {86}},\ \bibinfo {pages} {220401(R)}
  (\bibinfo {year} {2012})}\BibitemShut {NoStop}%
\bibitem [{\citenamefont {Maple}\ \emph {et~al.}(2010)\citenamefont {Maple},
  \citenamefont {Baumbach}, \citenamefont {Butch}, \citenamefont {Hamlin},\
  and\ \citenamefont {Janoschek}}]{MAPLE10}%
  \BibitemOpen
  \bibfield  {author} {\bibinfo {author} {\bibfnamefont {M.~B.}\ \bibnamefont
  {Maple}}, \bibinfo {author} {\bibfnamefont {R.~E.}\ \bibnamefont {Baumbach}},
  \bibinfo {author} {\bibfnamefont {N.~P.}\ \bibnamefont {Butch}}, \bibinfo
  {author} {\bibfnamefont {J.~J.}\ \bibnamefont {Hamlin}}, \ and\ \bibinfo
  {author} {\bibfnamefont {M.}~\bibnamefont {Janoschek}},\ }\href {\doibase
  10.1007/s10909-010-0212-5} {\bibfield  {journal} {\bibinfo  {journal} {J. Low
  Temp. Phys.}\ }\textbf {\bibinfo {volume} {161}},\ \bibinfo {pages} {4}
  (\bibinfo {year} {2010})}\BibitemShut {NoStop}%
\bibitem [{\citenamefont {{Gannon}}\ \emph {et~al.}(2017)\citenamefont
  {{Gannon}}, \citenamefont {{Wu}}, \citenamefont {{Zaliznyak}}, \citenamefont
  {{Xu}}, \citenamefont {{Tsvelik}}, \citenamefont {{Rodriguez-Rivera}},
  \citenamefont {{Qiu}},\ and\ \citenamefont {{Aronson}}}]{GANNON17u}%
  \BibitemOpen
  \bibfield  {author} {\bibinfo {author} {\bibfnamefont {W.~J.}\ \bibnamefont
  {{Gannon}}}, \bibinfo {author} {\bibfnamefont {L.~S.}\ \bibnamefont {{Wu}}},
  \bibinfo {author} {\bibfnamefont {I.~A.}\ \bibnamefont {{Zaliznyak}}},
  \bibinfo {author} {\bibfnamefont {W.}~\bibnamefont {{Xu}}}, \bibinfo {author}
  {\bibfnamefont {A.~M.}\ \bibnamefont {{Tsvelik}}}, \bibinfo {author}
  {\bibfnamefont {J.~A.}\ \bibnamefont {{Rodriguez-Rivera}}}, \bibinfo {author}
  {\bibfnamefont {Y.}~\bibnamefont {{Qiu}}}, \ and\ \bibinfo {author}
  {\bibfnamefont {M.~C.}\ \bibnamefont {{Aronson}}},\ }\href@noop {} {\enquote
  {\bibinfo {title} {{A Local Quantum Phase Transition in
  YFe$_{2}$Al$_{10}$}},}\ } (\bibinfo {year} {2017}),\ \Eprint
  {http://arxiv.org/abs/1712.04033} {arXiv:1712.04033 [cond-mat.str-el]}
  \BibitemShut {NoStop}%
\bibitem [{\citenamefont {Aji}\ and\ \citenamefont {Varma}(2010)}]{AJI10}%
  \BibitemOpen
  \bibfield  {author} {\bibinfo {author} {\bibfnamefont {V.}~\bibnamefont
  {Aji}}\ and\ \bibinfo {author} {\bibfnamefont {C.~M.}\ \bibnamefont
  {Varma}},\ }\href {\doibase 10.1103/PhysRevLett.82.174501} {\bibfield
  {journal} {\bibinfo  {journal} {Phys. Rev. B}\ }\textbf {\bibinfo {volume}
  {82}},\ \bibinfo {pages} {174501} (\bibinfo {year} {2010})}\BibitemShut
  {NoStop}%
\bibitem [{\citenamefont {Zhu}\ \emph {et~al.}(2015)\citenamefont {Zhu},
  \citenamefont {Chen},\ and\ \citenamefont {Varma}}]{ZHU15}%
  \BibitemOpen
  \bibfield  {author} {\bibinfo {author} {\bibfnamefont {L.}~\bibnamefont
  {Zhu}}, \bibinfo {author} {\bibfnamefont {Y.}~\bibnamefont {Chen}}, \ and\
  \bibinfo {author} {\bibfnamefont {C.~M.}\ \bibnamefont {Varma}},\ }\href
  {\doibase 10.1103/PhysRevB.91.205129} {\bibfield  {journal} {\bibinfo
  {journal} {Phys. Rev. B}\ }\textbf {\bibinfo {volume} {91}},\ \bibinfo
  {pages} {205129} (\bibinfo {year} {2015})}\BibitemShut {NoStop}%
\bibitem [{\citenamefont {Hou}\ and\ \citenamefont {Varma}(2016)}]{HOU16}%
  \BibitemOpen
  \bibfield  {author} {\bibinfo {author} {\bibfnamefont {C.}~\bibnamefont
  {Hou}}\ and\ \bibinfo {author} {\bibfnamefont {C.~M.}\ \bibnamefont
  {Varma}},\ }\href {\doibase 10.1103/PhysRevB.94.201101} {\bibfield  {journal}
  {\bibinfo  {journal} {Phys. Rev. B}\ }\textbf {\bibinfo {volume} {94}},\
  \bibinfo {pages} {201101} (\bibinfo {year} {2016})}\BibitemShut {NoStop}%
\bibitem [{\citenamefont {{Varma}}\ \emph {et~al.}(2017)\citenamefont
  {{Varma}}, \citenamefont {{Gannon}},\ and\ \citenamefont
  {{Aronson}}}]{VARMA17u}%
  \BibitemOpen
  \bibfield  {author} {\bibinfo {author} {\bibfnamefont {C.~M.}\ \bibnamefont
  {{Varma}}}, \bibinfo {author} {\bibfnamefont {W.~J.}\ \bibnamefont
  {{Gannon}}}, \ and\ \bibinfo {author} {\bibfnamefont {M.~C.}\ \bibnamefont
  {{Aronson}}},\ }\href@noop {} {\enquote {\bibinfo {title} {Quantum critical
  singularities in two-dimensional metallic {XY} ferromagnets},}\ } (\bibinfo
  {year} {2017}),\ \Eprint {http://arxiv.org/abs/1710.00380} {arXiv:1710.00380
  [cond-mat.str-el]} \BibitemShut {NoStop}%
\bibitem [{\citenamefont {Schenck}(1985)}]{SCHENCK85}%
  \BibitemOpen
  \bibfield  {author} {\bibinfo {author} {\bibfnamefont {A.}~\bibnamefont
  {Schenck}},\ }\href@noop {} {\emph {\bibinfo {title} {{Muon Spin Rotation
  Spectroscopy: Principles and Applications in Solid State Physics}}}}\
  (\bibinfo  {publisher} {A. Hilger},\ \bibinfo {address} {Bristol \& Boston},\
  \bibinfo {year} {1985})\BibitemShut {NoStop}%
\bibitem [{\citenamefont {Blundell}(1999)}]{BLUNDELL99}%
  \BibitemOpen
  \bibfield  {author} {\bibinfo {author} {\bibfnamefont {S.~J.}\ \bibnamefont
  {Blundell}},\ }\href {\doibase 10.1080/001075199181521} {\bibfield  {journal}
  {\bibinfo  {journal} {Contemp. Phys.}\ }\textbf {\bibinfo {volume} {40}},\
  \bibinfo {pages} {175} (\bibinfo {year} {1999})}\BibitemShut {NoStop}%
\bibitem [{\citenamefont {Brewer}(2003)}]{BREWER03}%
  \BibitemOpen
  \bibfield  {author} {\bibinfo {author} {\bibfnamefont {J.~H.}\ \bibnamefont
  {Brewer}},\ }in\ \href {\doibase 10.1002/3527600434.eap258} {\emph {\bibinfo
  {booktitle} {Digital Encyclopedia of Applied Physics}}},\ \bibinfo {editor}
  {edited by\ \bibinfo {editor} {\bibfnamefont {G.~L.}\ \bibnamefont {Trigg}},
  \bibinfo {editor} {\bibfnamefont {E.~S.}\ \bibnamefont {Vera}}, \ and\
  \bibinfo {editor} {\bibfnamefont {W.}~\bibnamefont {Greulich}}}\ (\bibinfo
  {publisher} {Wiley-VCH Verlag GmbH \& Co KGaA},\ \bibinfo {address}
  {Weinheim},\ \bibinfo {year} {2003})\BibitemShut {NoStop}%
\bibitem [{\citenamefont {Yaouanc}\ and\ \citenamefont {Dalmas~de
  R\'eotier}(2011)}]{YAOUANC11}%
  \BibitemOpen
  \bibfield  {author} {\bibinfo {author} {\bibfnamefont {A.}~\bibnamefont
  {Yaouanc}}\ and\ \bibinfo {author} {\bibfnamefont {P.}~\bibnamefont
  {Dalmas~de R\'eotier}},\ }\href@noop {} {\emph {\bibinfo {title} {Muon Spin
  Rotation, Relaxation, and Resonance: Applications to Condensed Matter}}},\
  {International Series of Monographs on Physics}\ (\bibinfo  {publisher}
  {Oxford University Press},\ \bibinfo {address} {New York},\ \bibinfo {year}
  {2011})\BibitemShut {NoStop}%
\bibitem [{\citenamefont {Moriya}(1963)}]{MORIYA63}%
  \BibitemOpen
  \bibfield  {author} {\bibinfo {author} {\bibfnamefont {T.}~\bibnamefont
  {Moriya}},\ }\href {\doibase 10.1143/JPSJ.18.516} {\bibfield  {journal}
  {\bibinfo  {journal} {J. Phys. Soc. Jpn.}\ }\textbf {\bibinfo {volume}
  {18}},\ \bibinfo {pages} {516} (\bibinfo {year} {1963})}\BibitemShut
  {NoStop}%
\bibitem [{\citenamefont {Jaccarino}(1967)}]{JACCARINO67}%
  \BibitemOpen
  \bibfield  {author} {\bibinfo {author} {\bibfnamefont {V.}~\bibnamefont
  {Jaccarino}},\ }in\ \href@noop {} {\emph {\bibinfo {booktitle} {Proceedings
  of the International School of Physics ``Enrico Fermi''}}},\ \bibinfo {series
  and number} {Course 37},\ \bibinfo {editor} {edited by\ \bibinfo {editor}
  {\bibfnamefont {W.}~\bibnamefont {Marshall}}}\ (\bibinfo  {publisher}
  {Academic Press},\ \bibinfo {address} {New York, London},\ \bibinfo {year}
  {1967})\ Chap.~\bibinfo {chapter} {5}, pp.\ \bibinfo {pages}
  {335--380}\BibitemShut {NoStop}%
\bibitem [{\citenamefont {Chakravarty}\ and\ \citenamefont
  {Orbach}(1990)}]{CHAKRAVARTY90}%
  \BibitemOpen
  \bibfield  {author} {\bibinfo {author} {\bibfnamefont {S.}~\bibnamefont
  {Chakravarty}}\ and\ \bibinfo {author} {\bibfnamefont {R.}~\bibnamefont
  {Orbach}},\ }\href@noop {} {\bibfield  {journal} {\bibinfo  {journal} {Phys.
  Rev. Lett.}\ }\textbf {\bibinfo {volume} {64}},\ \bibinfo {pages} {234}
  (\bibinfo {year} {1990})}\BibitemShut {NoStop}%
\bibitem [{\citenamefont {Adroja}\ \emph {et~al.}(2013)\citenamefont {Adroja},
  \citenamefont {Hillier}, \citenamefont {Muro}, \citenamefont {Takabatake},
  \citenamefont {Strydom}, \citenamefont {Bhattacharyya}, \citenamefont
  {Daoud-Aladin},\ and\ \citenamefont {Taylor}}]{ADROJA13}%
  \BibitemOpen
  \bibfield  {author} {\bibinfo {author} {\bibfnamefont {D.~T.}\ \bibnamefont
  {Adroja}}, \bibinfo {author} {\bibfnamefont {A.~D.}\ \bibnamefont {Hillier}},
  \bibinfo {author} {\bibfnamefont {Y.}~\bibnamefont {Muro}}, \bibinfo {author}
  {\bibfnamefont {T.}~\bibnamefont {Takabatake}}, \bibinfo {author}
  {\bibfnamefont {A.~M.}\ \bibnamefont {Strydom}}, \bibinfo {author}
  {\bibfnamefont {A.}~\bibnamefont {Bhattacharyya}}, \bibinfo {author}
  {\bibfnamefont {A.}~\bibnamefont {Daoud-Aladin}}, \ and\ \bibinfo {author}
  {\bibfnamefont {J.~W.}\ \bibnamefont {Taylor}},\ }\href@noop {} {\bibfield
  {journal} {\bibinfo  {journal} {Phys. Scr.}\ }\textbf {\bibinfo {volume}
  {88}},\ \bibinfo {pages} {068505} (\bibinfo {year} {2013})}\BibitemShut
  {NoStop}%
\bibitem [{\citenamefont {Kubo}\ and\ \citenamefont {Toyabe}(1967)}]{KUBO67}%
  \BibitemOpen
  \bibfield  {author} {\bibinfo {author} {\bibfnamefont {R.}~\bibnamefont
  {Kubo}}\ and\ \bibinfo {author} {\bibfnamefont {T.}~\bibnamefont {Toyabe}},\
  }in\ \href@noop {} {\emph {\bibinfo {booktitle} {{Magnetic Resonance and
  Relaxation}}}},\ \bibinfo {editor} {edited by\ \bibinfo {editor}
  {\bibfnamefont {R.}~\bibnamefont {Blinc}}}\ (\bibinfo  {publisher}
  {North-Holland},\ \bibinfo {address} {Amsterdam},\ \bibinfo {year} {1967})\
  pp.\ \bibinfo {pages} {810--823}\BibitemShut {NoStop}%
\bibitem [{\citenamefont {Hayano}\ \emph {et~al.}(1979)\citenamefont {Hayano},
  \citenamefont {Uemura}, \citenamefont {Imazato}, \citenamefont {Nishida},
  \citenamefont {Yamazaki},\ and\ \citenamefont {Kubo}}]{HAYANO79}%
  \BibitemOpen
  \bibfield  {author} {\bibinfo {author} {\bibfnamefont {R.~S.}\ \bibnamefont
  {Hayano}}, \bibinfo {author} {\bibfnamefont {Y.~J.}\ \bibnamefont {Uemura}},
  \bibinfo {author} {\bibfnamefont {J.}~\bibnamefont {Imazato}}, \bibinfo
  {author} {\bibfnamefont {N.}~\bibnamefont {Nishida}}, \bibinfo {author}
  {\bibfnamefont {T.}~\bibnamefont {Yamazaki}}, \ and\ \bibinfo {author}
  {\bibfnamefont {R.}~\bibnamefont {Kubo}},\ }\href@noop {} {\bibfield
  {journal} {\bibinfo  {journal} {Phys. Rev. B}\ }\textbf {\bibinfo {volume}
  {20}},\ \bibinfo {pages} {850} (\bibinfo {year} {1979})}\BibitemShut
  {NoStop}%
\bibitem [{\citenamefont {Adroja}\ \emph {et~al.}(2008)\citenamefont {Adroja},
  \citenamefont {Hillier}, \citenamefont {Park}, \citenamefont {Kockelmann},
  \citenamefont {McEwen}, \citenamefont {Rainford}, \citenamefont {Jang},
  \citenamefont {Geibel},\ and\ \citenamefont {Takabatake}}]{ADROJA08}%
  \BibitemOpen
  \bibfield  {author} {\bibinfo {author} {\bibfnamefont {D.~T.}\ \bibnamefont
  {Adroja}}, \bibinfo {author} {\bibfnamefont {A.~D.}\ \bibnamefont {Hillier}},
  \bibinfo {author} {\bibfnamefont {J.-G.}\ \bibnamefont {Park}}, \bibinfo
  {author} {\bibfnamefont {W.}~\bibnamefont {Kockelmann}}, \bibinfo {author}
  {\bibfnamefont {K.~A.}\ \bibnamefont {McEwen}}, \bibinfo {author}
  {\bibfnamefont {B.~D.}\ \bibnamefont {Rainford}}, \bibinfo {author}
  {\bibfnamefont {K.-H.}\ \bibnamefont {Jang}}, \bibinfo {author}
  {\bibfnamefont {C.}~\bibnamefont {Geibel}}, \ and\ \bibinfo {author}
  {\bibfnamefont {T.}~\bibnamefont {Takabatake}},\ }\href {\doibase
  10.1103/PhysRevB.78.014412} {\bibfield  {journal} {\bibinfo  {journal} {Phys.
  Rev. B}\ }\textbf {\bibinfo {volume} {78}},\ \bibinfo {pages} {014412}
  (\bibinfo {year} {2008})}\BibitemShut {NoStop}%
\bibitem [{\citenamefont {Carretta}\ \emph {et~al.}(2009)\citenamefont
  {Carretta}, \citenamefont {Pasero}, \citenamefont {Giovannini},\ and\
  \citenamefont {Baines}}]{CARRETTA09}%
  \BibitemOpen
  \bibfield  {author} {\bibinfo {author} {\bibfnamefont {P.}~\bibnamefont
  {Carretta}}, \bibinfo {author} {\bibfnamefont {R.}~\bibnamefont {Pasero}},
  \bibinfo {author} {\bibfnamefont {M.}~\bibnamefont {Giovannini}}, \ and\
  \bibinfo {author} {\bibfnamefont {C.}~\bibnamefont {Baines}},\ }\href
  {\doibase 10.1103/PhysRevB.79.020401} {\bibfield  {journal} {\bibinfo
  {journal} {Phys. Rev. B}\ }\textbf {\bibinfo {volume} {79}},\ \bibinfo
  {pages} {020401} (\bibinfo {year} {2009})}\BibitemShut {NoStop}%
\bibitem [{\citenamefont {Spehling}\ \emph {et~al.}(2012)\citenamefont
  {Spehling}, \citenamefont {G{\"u}nther}, \citenamefont {Krellner},
  \citenamefont {Y\`{e}che}, \citenamefont {Leutkens}, \citenamefont {Baines},
  \citenamefont {Geibel},\ and\ \citenamefont {Klauss}}]{SPEHLING12}%
  \BibitemOpen
  \bibfield  {author} {\bibinfo {author} {\bibfnamefont {J.}~\bibnamefont
  {Spehling}}, \bibinfo {author} {\bibfnamefont {M.}~\bibnamefont
  {G{\"u}nther}}, \bibinfo {author} {\bibfnamefont {C.}~\bibnamefont
  {Krellner}}, \bibinfo {author} {\bibfnamefont {N.}~\bibnamefont {Y\`{e}che}},
  \bibinfo {author} {\bibfnamefont {H.}~\bibnamefont {Leutkens}}, \bibinfo
  {author} {\bibfnamefont {C.}~\bibnamefont {Baines}}, \bibinfo {author}
  {\bibfnamefont {C.}~\bibnamefont {Geibel}}, \ and\ \bibinfo {author}
  {\bibfnamefont {H.-H.}\ \bibnamefont {Klauss}},\ }\href {\doibase
  10.1103/PhysRevB.85.140406} {\bibfield  {journal} {\bibinfo  {journal} {Phys.
  Rev. B}\ }\textbf {\bibinfo {volume} {85}},\ \bibinfo {pages} {140406}
  (\bibinfo {year} {2012})}\BibitemShut {NoStop}%
\bibitem [{\citenamefont {Sarkar}\ \emph {et~al.}(2017)\citenamefont {Sarkar},
  \citenamefont {Spehling}, \citenamefont {Materne}, \citenamefont {Luetkens},
  \citenamefont {Baines}, \citenamefont {Brando}, \citenamefont {Krellner},\
  and\ \citenamefont {Klauss}}]{SARKAR17}%
  \BibitemOpen
  \bibfield  {author} {\bibinfo {author} {\bibfnamefont {R.}~\bibnamefont
  {Sarkar}}, \bibinfo {author} {\bibfnamefont {J.}~\bibnamefont {Spehling}},
  \bibinfo {author} {\bibfnamefont {P.}~\bibnamefont {Materne}}, \bibinfo
  {author} {\bibfnamefont {H.}~\bibnamefont {Luetkens}}, \bibinfo {author}
  {\bibfnamefont {C.}~\bibnamefont {Baines}}, \bibinfo {author} {\bibfnamefont
  {M.}~\bibnamefont {Brando}}, \bibinfo {author} {\bibfnamefont
  {C.}~\bibnamefont {Krellner}}, \ and\ \bibinfo {author} {\bibfnamefont
  {H.-H.}\ \bibnamefont {Klauss}},\ }\href {\doibase
  10.1103/PhysRevB.95.121111} {\bibfield  {journal} {\bibinfo  {journal} {Phys.
  Rev. B}\ }\textbf {\bibinfo {volume} {95}},\ \bibinfo {pages} {121111}
  (\bibinfo {year} {2017})}\BibitemShut {NoStop}%
\bibitem [{Note1()}]{Note1}%
  \BibitemOpen
  \bibinfo {note} {In zero applied field the muon Zeeman splitting is due to
  nuclear dipolar fields, which are randomly oriented and tend to average out
  any anisotropy.}\BibitemShut {Stop}%
\bibitem [{\citenamefont {Maisuradze}\ \emph {et~al.}(2010)\citenamefont
  {Maisuradze}, \citenamefont {Schnelle}, \citenamefont {Khasanov},
  \citenamefont {Gumeniuk}, \citenamefont {Nicklas}, \citenamefont {Rosner},
  \citenamefont {Leithe-Jasper}, \citenamefont {Grin}, \citenamefont {Amato},\
  and\ \citenamefont {Thalmeier}}]{MAISURADZE10}%
  \BibitemOpen
  \bibfield  {author} {\bibinfo {author} {\bibfnamefont {A.}~\bibnamefont
  {Maisuradze}}, \bibinfo {author} {\bibfnamefont {W.}~\bibnamefont
  {Schnelle}}, \bibinfo {author} {\bibfnamefont {R.}~\bibnamefont {Khasanov}},
  \bibinfo {author} {\bibfnamefont {R.}~\bibnamefont {Gumeniuk}}, \bibinfo
  {author} {\bibfnamefont {M.}~\bibnamefont {Nicklas}}, \bibinfo {author}
  {\bibfnamefont {H.}~\bibnamefont {Rosner}}, \bibinfo {author} {\bibfnamefont
  {A.}~\bibnamefont {Leithe-Jasper}}, \bibinfo {author} {\bibfnamefont
  {Y.}~\bibnamefont {Grin}}, \bibinfo {author} {\bibfnamefont {A.}~\bibnamefont
  {Amato}}, \ and\ \bibinfo {author} {\bibfnamefont {P.}~\bibnamefont
  {Thalmeier}},\ }\href {\doibase 10.1103/PhysRevB.82.024524} {\bibfield
  {journal} {\bibinfo  {journal} {Phys. Rev. B}\ }\textbf {\bibinfo {volume}
  {82}},\ \bibinfo {pages} {024524} (\bibinfo {year} {2010})}\BibitemShut
  {NoStop}%
\bibitem [{\citenamefont {Keren}\ \emph {et~al.}(1996)\citenamefont {Keren},
  \citenamefont {Mendels}, \citenamefont {Campbell},\ and\ \citenamefont
  {Lord}}]{KEREN96}%
  \BibitemOpen
  \bibfield  {author} {\bibinfo {author} {\bibfnamefont {A.}~\bibnamefont
  {Keren}}, \bibinfo {author} {\bibfnamefont {P.}~\bibnamefont {Mendels}},
  \bibinfo {author} {\bibfnamefont {I.~A.}\ \bibnamefont {Campbell}}, \ and\
  \bibinfo {author} {\bibfnamefont {J.}~\bibnamefont {Lord}},\ }\href {\doibase
  10.1103/PhysRevLett.77.1386} {\bibfield  {journal} {\bibinfo  {journal}
  {Phys. Rev. Lett.}\ }\textbf {\bibinfo {volume} {77}},\ \bibinfo {pages}
  {1386} (\bibinfo {year} {1996})}\BibitemShut {NoStop}%
\bibitem [{\citenamefont {Keren}\ \emph {et~al.}(2001)\citenamefont {Keren},
  \citenamefont {Bazalitsky}, \citenamefont {Campbell},\ and\ \citenamefont
  {Lord}}]{KEREN01}%
  \BibitemOpen
  \bibfield  {author} {\bibinfo {author} {\bibfnamefont {A.}~\bibnamefont
  {Keren}}, \bibinfo {author} {\bibfnamefont {G.}~\bibnamefont {Bazalitsky}},
  \bibinfo {author} {\bibfnamefont {I.}~\bibnamefont {Campbell}}, \ and\
  \bibinfo {author} {\bibfnamefont {J.~S.}\ \bibnamefont {Lord}},\ }\href
  {\doibase 10.1103/PhysRevB.64.054403} {\bibfield  {journal} {\bibinfo
  {journal} {Phys. Rev. B}\ }\textbf {\bibinfo {volume} {64}},\ \bibinfo
  {pages} {054403} (\bibinfo {year} {2001})}\BibitemShut {NoStop}%
\bibitem [{\citenamefont {Adam}\ \emph {et~al.}(2014)\citenamefont {Adam},
  \citenamefont {Suprayoga}, \citenamefont {Adiperdana}, \citenamefont {Guo},
  \citenamefont {Tanida}, \citenamefont {Mohd-Tajudin}, \citenamefont
  {Kobayashi}, \citenamefont {Sera}, \citenamefont {Nishioka}, \citenamefont
  {Matsumura}, \citenamefont {Sulaiman}, \citenamefont {Mohamed-Ibrahim},\ and\
  \citenamefont {Watanabe}}]{ADAM14}%
  \BibitemOpen
  \bibfield  {author} {\bibinfo {author} {\bibfnamefont {N.}~\bibnamefont
  {Adam}}, \bibinfo {author} {\bibfnamefont {E.}~\bibnamefont {Suprayoga}},
  \bibinfo {author} {\bibfnamefont {B.}~\bibnamefont {Adiperdana}}, \bibinfo
  {author} {\bibfnamefont {H.}~\bibnamefont {Guo}}, \bibinfo {author}
  {\bibfnamefont {H.}~\bibnamefont {Tanida}}, \bibinfo {author} {\bibfnamefont
  {S.~S.}\ \bibnamefont {Mohd-Tajudin}}, \bibinfo {author} {\bibfnamefont
  {R.}~\bibnamefont {Kobayashi}}, \bibinfo {author} {\bibfnamefont
  {M.}~\bibnamefont {Sera}}, \bibinfo {author} {\bibfnamefont {T.}~\bibnamefont
  {Nishioka}}, \bibinfo {author} {\bibfnamefont {M.}~\bibnamefont {Matsumura}},
  \bibinfo {author} {\bibfnamefont {S.}~\bibnamefont {Sulaiman}}, \bibinfo
  {author} {\bibfnamefont {M.~I.}\ \bibnamefont {Mohamed-Ibrahim}}, \ and\
  \bibinfo {author} {\bibfnamefont {I.}~\bibnamefont {Watanabe}},\ }\href
  {http://stacks.iop.org/1742-6596/551/i=1/a=012053} {\bibfield  {journal}
  {\bibinfo  {journal} {J. Phys.: Conf. Ser.}\ }\textbf {\bibinfo {volume}
  {551}},\ \bibinfo {pages} {012053} (\bibinfo {year} {2014})}\BibitemShut
  {NoStop}%
\bibitem [{\citenamefont {Kirkpatrick}\ and\ \citenamefont
  {Belitz}(2014)}]{KIRKPATRICK14}%
  \BibitemOpen
  \bibfield  {author} {\bibinfo {author} {\bibfnamefont {T.~R.}\ \bibnamefont
  {Kirkpatrick}}\ and\ \bibinfo {author} {\bibfnamefont {D.}~\bibnamefont
  {Belitz}},\ }\href {\doibase 10.1103/PhysRevLett.113.127203} {\bibfield
  {journal} {\bibinfo  {journal} {Phys. Rev. Lett.}\ }\textbf {\bibinfo
  {volume} {113}},\ \bibinfo {pages} {127203} (\bibinfo {year}
  {2014})}\BibitemShut {NoStop}%
\end{thebibliography}

\end{document}